\documentclass[12pt,twoside,dvips]{article}
\usepackage{epsfig}
\usepackage{graphics}
\usepackage{a4wide}
\def\fmn#1#2{\mbox{${\textstyle \frac{#1}{#2}}$}}

\begin{document}
\mbox{ }
\vspace{20mm}

\includegraphics{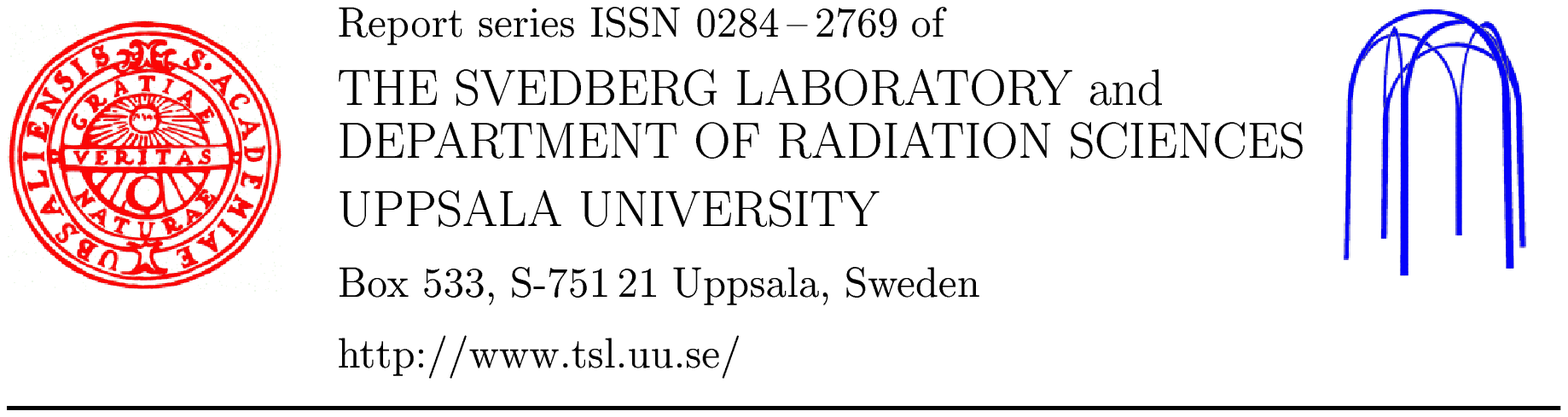}

\begin{flushright}
\begin{minipage}[t]{37mm}
{\bf TSL/ISV-98-0198 \\
October 1998} 
\end{minipage}
\end{flushright}

\begin{center}
\vspace{5mm}

{\bf \Large Higher Partial Waves in
\mbox{\boldmath\large $pp\to pp\eta$} near Threshold }

\vspace{10mm}

{ 
H.~Cal\'en~$^a$,
J.~Dyring~$^a$,
G.~F\"aldt~$^a$,
K.~Fransson~$^a$,
L.~Gustafsson~$^a$,
S.~H\"aggstr\"om~$^a$,
B.~H\"oistad~$^a$,
J.~Johanson~$^a$,
A.~Johansson~$^a$,
T.~Johansson~$^a$,
S.~Kullander~$^a$,
A.~M\"ortsell~$^a$,
R.~Ruber~$^a$,
J.~Z\l oma\'nczuk~$^a$,
C.~Ekstr\"om~$^b$,
K.~Kilian~$^c$,
W.~Oelert~$^c$,
T.~Sefzick~$^c$,
R.~Bilger~$^d$,
W.~Brodowski~$^d$,
H.~Clement~$^d$,
G.J.~Wagner~$^d$,
A.~Bondar~$^e$,
A.~Kuzmin~$^e$,
B.~Shwartz~$^e$,
V.~Sidorov~$^e$,
A.~Sukhanov~$^e$,
V.~Dunin~$^f$,
B.~Morosov~$^f$,
A.~Povtorejko~$^f$,
A.~Sukhanov~$^f$,
A.~Zernov~$^f$,
A.~Kupsc~$^g$,
P.~Marciniewski~$^g$,
J.~Stepaniak~$^g$,
J.~Zabierowski~$^h$,
A.~Turowiecki~$^i$,
Z.~Wilhelmi~$^i$,
C.~Wilkin~$^j$
 \\ }
\vspace{3mm}
$^a$ Department of Radiation Science, Uppsala University, S-751 21 Uppsala,
Sweden\\
$^b$ The Svedberg Laboratory, S-751 21 Uppsala, Sweden\\
$^c$ IKP, Forschungszentrum J\"ulich GmbH, D-52425 J\"ulich, Germany\\
$^d$ Physikalisches Institut, T\"ubingen University, D-72076 T\"ubingen,
Germany \\
$^e$ Institute of Nuclear Physics, Novosibirsk 630 090, Russia\\
$^f$ Joint Institute for Nuclear Research Dubna, 101000 Moscow, Russia\\
$^g$ Institute for Nuclear Studies, PL-00681 Warsaw, Poland\\
$^h$ Institute for Nuclear Studies, PL-90137 L\'odz, Poland\\
$^i$ Institute of Experimental Physics, Warsaw University, PL-0061 Warsaw,
Poland\\
$^j$ Physics \& Astronomy Dept., University College London,
London WC1E 6BT, U.K.\\
\end{center}

\vspace{5mm}

{\bf Abstract:}
Exclusive measurements of the production of $\eta$-mesons in the $pp\to pp\eta$
reaction have been carried out at excess energies of 16 and 37~MeV above
threshold. The deviations from phase space are dominated by the proton-proton
final state
interaction and this influences particularly the energy distribution of the
$\eta$ meson. However, evidence is also presented at the higher energy for the
existence of an anisotropy in the angular distributions of the $\eta$-meson and
also of the final proton-proton pair, probably to be associated with $D$-waves
in this system interfering with the dominant $S$-wave term. The sign of the
$\eta$ angular anisotropy suggests that $\rho$-exchange is important for this
reaction.\\

\vfill
\clearpage
\setcounter{page}{1}

\baselineskip 4ex

The production of $\eta$ mesons in proton-proton collisions near threshold has
been measured in recent years by three different
groups~\cite{SPESIII,PINOT,TSL1} and a fairly consistent picture has emerged. In
comparison to the analogous pion case~\cite{IUCF,TSL2}, $S$-wave production is
more dominant near threshold, and this is generally ascribed to the presence of
the N$^*$(1535) S$_{11}$ isobar, whose width overlaps the $\eta-p$ threshold and
which has large branching fractions into both $\eta-p$ and $\pi-p$~\cite{PDG}.
Dalitz plots obtained for the $pp\to pp\eta$ reaction at low energy show strong
deviations from phase space due to the presence of the proton-proton final
state interaction (FSI)~\cite{TSL1}. In addition there are residual effects but
the precision of these measurements was limited, in part, by the uncertainty in
the determination of the proton angles. This defect has been overcome in the
present experiment through the inclusion of a tracking device covering the
forward angles and operating in conjunction with the apparatus previously used.
The combination allowed us to investigate the influence of higher partial waves
in the angular distributions and to identify dependences on both the angles of
the $\eta$ and of the proton-proton relative momentum, which become stronger
with energy. This behaviour is the first indication of effects from higher
partial waves in the $pp\to pp\eta$ reaction. The shape of the $\eta$ angular
variation is sensitive to the basic production mechanism and the data suggest
that $\rho$-exchange provides a more important driving term here than
$\pi$-exchange. 

The experiment was carried out using the PROMICE/WASA facility at the CELSIUS
storage ring of the The Svedberg Laboratory at beam energies of 1296 and
1350~MeV, corresponding to centre-of-mass excess energies $Q = 16$ and
37~MeV respectively. Using a cluster gas jet target, integrated luminosities
of about 50~nb$^{-1}$ and 200~nb$^{-1}$ were obtained for the two energies 
in a total of 30 hours of running, and these yielded about 300 and 750 good
$pp\eta$ events in the final sample. Details of the detector system are given
in Refs.~\cite{PROMICE,Dyring}, and only the main points are discussed here. 

The forward-going protons were measured in a detector system covering polar
angles between 4$^{\circ}$ and 20$^{\circ}$ with respect to the beam direction.
It consists of a tracking detector, followed by a three-layer scintillator
hodoscope and a four-layer scintillator calorimeter. A second hodoscope is
placed at the end of the detector system to register penetrating particles. The
tracker consists of two planes, each with four layers of thin-walled cylindrical
drift chambers, so-called ``straw chambers'', oriented in the vertical and
horizontal directions. This arrangement allows the proton scattering angles to
be reconstructed to a precision of better than 1$^{\circ}$ (FWHM). The $\eta$'s
are identified from their 2$\gamma$ decay channel, where the $\gamma$'s are
detected in two CsI(Na) arrays situated on either side of the scattering
chamber. Scintillator hodoscopes are placed in front of the CsI arrays to veto
charged particles. The 2$\gamma$ invariant mass resolution obtained at the
$\eta$-meson mass is 20 MeV (RMS).

In the off-line analysis, only those events with an identified $\eta$ meson
together with two energetic protons in the forward detector were retained and,
in these cases, a kinematical fit was applied in order to extract the
energy-momentum vectors for the particles. To avoid problems in accounting for
losses through nuclear reactions in the detector material, only the proton
directions were used, and this results in a fit with three constraints (3C). The
number of background events with two uncorrelated $\gamma$'s arising from
different pions in 2$\pi^0$ production becomes negligible after applying a
lower cut at 5\% on the confidence level of the fit. In the subsequent analysis
of our data, we exploit the fact that at low energies only a few amplitudes are
allowed and that these lead to distributions in which the angular and momentum
variables are intimately linked. It is only through the introduction of such a
functional form into Monte Carlo simulations that we can draw any firm
conclusions on the physics of the process. The detector response and acceptance
calculations for the experiment were made with events generated from a full
Monte Carlo simulation using GEANT3~\cite{Geant}. The resulting geometrical
acceptance for this type of event is about 0.5\% and 2\% at $Q = 16$ and 37~MeV
respectively. The data were checked for internal consistency by applying
different geometrical cuts.

The sole production amplitude which survives at threshold corresponds to the
transition $^{3\!}P_0\to\, ^{1\!}S_0\,s$, where we are using the standard
$^{2S+1\!}L_J$ notation for the $pp$ system, with the lower case letter
denoting the angular momentum of the $\eta$-meson with respect to the $pp$
system. At slightly higher energies, amplitudes corresponding to the production
of $P$- and $D$-wave $pp$ pairs introduce dependences on the $pp$ angles with
similar momentum threshold factors since the $D$-wave term can interfere with
the threshold $S$-wave amplitude. However, given that $\eta$ production in the
$^{1\!}S_0\,p$ state is forbidden by selection rules, the first non-trivial 
$\eta$ angular dependence is expected to come from the interference of the 
$s$- and $d$-wave amplitudes. These considerations lead us to take the 
following simple form for the low energy $pp\to pp\eta$ amplitude:
\begin{equation}
M=\tilde{A}_{Ss}\:\phi_{f}^{\dagger}\,(\hat{p}\cdot\vec{\varepsilon}_{i})
+A_{Sd}\:\phi_{f}^{\dagger}\,
(\hat{p}\cdot\vec{k})(\vec{k}\cdot\vec{\varepsilon}_{i})
+A_{Ps}\:\phi_i\,(\vec{q}\cdot\vec{\varepsilon}_f^{\,\dagger})
+A_{Ds}\:\phi_f^{\dagger}\,
(\hat{p}\cdot\vec{q})(\vec{q}\cdot\vec{\varepsilon}_i)\:.
\end{equation}
The momenta of the
initial proton and final $\eta$ in the overall c.m.\ system are denoted by
$\vec{p}$ and $\vec{k}$ respectively, $2\vec{q}$ is the relative momentum in
the final two-proton system, $\vec{\varepsilon}_i$~($\vec{\varepsilon}_f$) the
spin-one polarisation vector of the initial (final) proton-proton pair, and
$\phi_i$~($\phi_f$) the corresponding spin-zero functions. 
The indices on the amplitudes denote the dominant final angular momentum
state in a term. It should however be noted that the strict $Ss$ partial wave
amplitude is given by
\begin{equation}
\label{Relation}
A_{Ss} = \tilde{A}_{Ss} +\fmn{1}{3}\,k^2\,A_{Sd} + \fmn{1}{3}\,q^2\,A_{Ds}\:.
\end{equation}

Keeping only terms up to order $k^2$ or $q^2$, {\it i.e.}\ a total of two units
of angular momentum in the final intensity, the spin-averaged matrix
element squared becomes
\begin{equation}
\label{M-squared}
\overline{|M|^2}= \frac{1}{4}\left[|\tilde{A}_{Ss}|^2 +
2\,k^2\,Re\left\{\tilde{A}_{Ss}^*A_{Sd}\right\}\,\cos^2\theta_{\eta} +
2\,q^2\,Re\left\{\tilde{A}_{Ss}^*A_{Ds}\right\}\,\cos^2\theta_{pp} 
+q^2\,|A_{Ps}|^2\right]\:,
\end{equation}
where $\theta_{\eta}$ and $\theta_{pp}$ are the angles that the $\eta$ and the 
$pp$ relative momentum make with respect to the beam direction. 

In a model where the N$^*(1535)$ isobar is excited through one pion
exchange~\cite{GW}, it is expected that the $|A_{Ps}|^2$ term should be smaller
than the $Re\left\{\tilde{A}_{Ss}^*A_{Ds}\right\}$ of the $Ss$-$Ds$ interference
by a factor of the order of $\mu/4m_p$, where $m_{\eta}$ and $m_p$ are the the
$\eta$ and proton masses respectively, and $\mu = 2m_pm_{\eta}/(2m_p+m_{\eta})$
is the reduced mass of the final state. Since the $Ps$ term has no
characteristic angular dependence, it would be difficult to isolate a small
effect as compared to a possible energy dependence of $\tilde{A}_{Ss}$, and so 
any such contribution is ignored.

The $^{1\!}S_0$ state of the final $pp$ system is subject to a very strong FSI
which is central to any analysis of low energy production. All the terms in
Eq.~(\ref{M-squared}) are influenced by the FSI, with the exception of the
$P$-wave contribution, and this will reduce even further the relative importance
of $|A_{Ps}|^2$. Without knowing the radial dependence of the $\eta$-production
operator, the FSI effect is slightly model-dependent. We estimate it by taking
the ratio of the Paris $^{1\!}S_0$ wave function~\cite{Paris} at its maximum at
$r=1$~fm to the corresponding plane wave function; numerical calculations with
the Paris wave functions show that Coulomb effects are negligible under the
conditions of the present experiment. There is a common enhancement factor
$F_{SS}(q)$ for the $(Ss)^2$ and $Ss$-$Sd$ interference terms and an analogous
$F_{SD}(q)$ for the $Ss$-$Ds$ interference. They may be parametrised in a
similar manner to that given in ref.~\cite{FW}
\begin{eqnarray}
\nonumber
F_{SS}(q)&=&0.440+\frac{151.7}{1+q^2/\alpha^2}\:,\\ 
F_{SD}(q)&=&0.968+\frac{11.5}{1+q^2/\alpha^2} \:,
\end{eqnarray}
with $\alpha=-0.053$~fm$^{-1}$.

We make the assumption that, apart from the FSI, the amplitudes
$\tilde{A}_{Ss}$, $A_{Sd}$ and $A_{Ds}$ appearing in Eq.~(\ref{M-squared}) are
constant. The resulting form of the differential cross section is
\begin{equation}
\label{sigma}
d\sigma = \frac{N}{p}\:\left(F_{SS}(q)
+ a\,\frac{k^2}{\mu m_p}\:F_{SS}(q) \cos^2\theta_{\eta}
+ b\,\frac{q^2}{\mu m_p}\: F_{SD}(q) \cos^2\theta_{pp}\right) 
\mbox{\it dLips}\:,
\end{equation}
where $N$ is a normalization constant and {\it dLips} is the invariant
three-body phase space distribution. Typical momentum factors $\mu m_p$ are
explicitly shown to leave two dimensionless constants $a$ and $b$ to be
determined from the shapes of the angular and energy distributions. It must be 
stressed that the functional form of Eq.~(\ref{sigma}) does not depend
upon the details of a specific dynamical model but on the assumption of
constant amplitudes.

To extract differential distributions from the data, we fitted the Monte Carlo
simulations, with events weighted according to Eq.~(4), to the experimental
results. Following this procedure at $Q = 37$~MeV, a combined fit to the
$\theta_{\eta}$ and $\theta_{pp}$ angular distributions yields the parameter
values $a= -15 \pm 5$ and $b= 22\pm 7$, which are correlated, and
the overall normalisation constant. In addition to the given
statistical errors, there are systematic errors, arising mainly from
uncertainties in the geometrical alignment of the apparatus, which are of the
order 10\% for the $a$ parameter and 20\% for the $b$ parameter. These values
were then used in the Monte Carlo simulation to make the acceptance corrections
needed to unfold the data; the resulting corrected experimental data and
fitted angular distributions are shown in Figs.~1a and 2a. It should be noted
that the $a$ and $b$ parameters were not deduced by fitting to the corrected
data shown in the figures. 

The distribution in the $\eta$ kinetic energy $T_{\eta}$, shown in Fig.~3a, is
shifted towards higher energies with respect to phase space. This is a direct
consequence of the strongly attractive $pp$ FSI, which enhances events where the
$\eta$ recoils against the two protons which have low excitation energy. The
dashed curve, which corresponds to phase space modified by the FSI, produces too
big an effect. The solid curve is the prediction with the values of $a$ and $b$
as determined by the angular distributions. The non-vanishing of the averages of
$\cos^2\theta$ leaves $k^2$ and $q^2$ terms in Eq.~(\ref{sigma}) which yield a
much better fit. 

At $Q =16$~MeV, S-waves are even more dominant and the angular distributions are
broadly compatible with phase space modified by the pp FSI. Nevertheless the
shapes are in fact marginally better reproduced with the parameter values
derived from the higher energy data and the corresponding predictions are shown
in Figs.~1b and 2b. The $T_{\eta}$ data shown in Fig.~3b, while again
demonstrating the effect of the $pp$ FSI, are well described by the
parametrisation. The experimental points do depend, to some extent, on the shape
of the assumed differential distribution, but this is well within the
statistical uncertainty of the data. The poor quality of the data for
$\cos\theta_{pp} > 0.8$ and at the upper end of the $T_{\eta}$ distribution
reflects mainly the loss of events due to the beam pipe so that discrepancies
with the parametrisation should not be taken seriously in these regions.

The numerical values of the differential cross sections are given in
Tables~1--3. The data were normalised to the
values of the total $pp\to pp\eta$ cross sections given in Ref.~\cite{TSL1},
namely $\sigma(T_p=1293~\mbox{\rm MeV}) = (2.11\pm 0.32)~\mu$b and 
$\sigma(T_p=1352~\mbox{\rm MeV}) = (4.92\pm 0.74)~\mu$b. 

All existing theoretical models describing $\eta$-production in proton-proton
scattering are broadly similar~\cite{GW,Theory}, consisting of a meson exchange
exciting a nucleon isobar, dominantly the $S_{11}$ N$^*$(1535), which decays
into an $s$-wave $\eta$-proton pair. The models differ mainly in their choice 
of mesons exchanged and, in particular, the relative importance of $\pi$ and 
$\rho$ exchange~\cite{GW,Theory}. If, for simplicity, only the pion exchange 
term is retained then it follows from the expansion of the corresponding 
propagator and vertex function that $a=1$ and $b=4$~\cite{FW7}. Our experimental
values are significantly larger than these.

Higher partial wave N$^*$ resonances are potentially very important for the 
$\eta$ angular distribution. Away from threshold there is evidence for
significant $d$-wave production of the $\eta$ meson in the $\pi^-p\to \eta n$
reaction~\cite{Deinet} and this is confirmed in the preliminary high statistics
data from the Crystal Ball collaboration~\cite{Ben}. Whilst having no
appreciable effect on the $pp$ angular distribution, the inclusion of such 
$d$-wave production in a one pion exchange model would contribute about 
$a\approx +7$ in the $\eta$ angular distribution which, though of the right 
order of magnitude, is of opposite sign to what we have deduced from our data. 
Thus, in contrast to our findings presented in Fig.~1b, such a term would 
favour production towards $\cos\theta_{\eta} =\pm 1$. 

However the sign of the $\cos^2\theta_{\eta}$ term is negative in the
photoproduction $\gamma p\to\eta p$ a little above threshold~\cite{Krusche}
and, using vector dominance ideas, this is likely to be true for $\rho p\to\eta
p$ as well. In a pure $\rho$-exchange model, the elementary distribution would
contribute $a\approx -2.5$ which, though too small, is of the same sign as the
one apparent in our data. In the original estimation of this process~\cite{GW}
it was claimed that $\rho$-exchange should be stronger than $\pi$-exchange and
that the interference between them was mainly destructive. If this were indeed
the case, the $a$ coefficient could be enhanced significantly because of the
negative sign between the $\rho$ and $\pi$ amplitudes. Our data would support
such a conclusion. This would also lead to the prediction that the $pn\to
d\eta$ should show a much flatter distribution since the $\rho$ and $\pi$
amplitudes add in this case~\cite{GW}. 

In conclusion, we have presented the first experimental evidence for
non-isotropy in the $pp \to pp\eta$ angular distributions close to threshold.
The signals are generally small in the data as compared for example to the
proton-proton final state interaction, which has overwhelming importance.
Nevertheless, a clear sign of an $\eta$ angular anisotropy has been found which
could be the first direct indication of $\rho$-dominance in $\eta$ production.

An economic description of all our distributions, sufficient for acceptance
corrections, has been given in terms of the two free parameters of 
Eq.~(\ref{sigma}) by taking the amplitudes $\tilde{A}_{Ss}$, $A_{Sd}$ 
and $A_{Ds}$ to be constant. If, for example, we assume instead that 
$A_{Ss}$ of Eq.~(\ref{Relation}) is constant then the $\cos^2\theta$'s in 
Eq.~(\ref{sigma}) are replaced by Legendre polynomials $P_2(\cos\theta)$. 
The fits to the angular distributions in Figs.~1 and 2 are very similar
but, due to the vanishing of the angular average of $P_2(\cos\theta)$, the
results in Fig.~3 are identical to the broken curve, which represents just 
phase space and FSI. To restore the previous good agreement requires an 
additional free $q^2$ or $k^2$ term in the fitting and, when this is introduced,
the results are essentially identical to the solid curve of Fig.~3. The shapes
of the angular distributions are unaffected by such a modified procedure.

A significant improvement of the statistics would be welcome to tie down the
model parameters in this area. Such an improvement is anticipated through the
use of the WASA detector~\cite{WASA}, which is designed for the study of rare
decays of the $\eta$ meson and which will have an almost $4\pi$ coverage of the
photons from the $\eta$ decay. The effects of higher partial waves should
increase strongly with beam energy but to exploit this would, in our case,
require an energy upgrade of CELSIUS or the study of quasi-free production on
the deuteron. 

We are grateful to the TSL/ISV personnel for their continued help during
the course of this work. Discussions with M.~Gar\c{c}on on the analysis of this
experiment were very helpful. Financial support for this experiment and its
analysis was provided by the Swedish Natural Science Research Council, the
Swedish Royal Academy of Science, the Swedish Institute, the Bundesministerium
f\"ur Bildung und Forschung (06TU886), Deutsche Forschung Gesellschaft (Mu
705/3 Graduiertenkolleg), the Polish State Committee for Scientific Research,
the Russian Academy of Science, and the European Science Exchange Programme.
The data presented here are based upon the work of J.~Dyring, in partial
fulfillment of the Ph.D.\ requirements ~\cite{Dyring}.

\newpage

\begin{table}[hp]
\caption[table1]{Differential cross section with respect to the $\eta$ c.m.\
angle for the $pp\to pp\eta$ reaction at $Q=16$ and 37~MeV ($T_p=1296$ and
1350~MeV). In addition to the statistical error, there is an overall systematic
uncertainty of about 20\%.}
\centering
\vspace{2ex}
\begin{tabular}{|c|c||c|c|}
\hline
\multicolumn{2}{|c||}{}&
\multicolumn{2}{c|}{}\\
\multicolumn{2}{|c||}{$Q =16$~MeV}&
\multicolumn{2}{c|}{$Q =37$~MeV}\\
\multicolumn{2}{|c||}{}&
\multicolumn{2}{c|}{}\\
\hline
&&&\\
$\cos\theta_{\eta}$&$d\sigma/d\Omega_{\eta}$ ($\mu$b/sr)&
$\cos\theta_{\eta}$&$d\sigma/d\Omega_{\eta}$ ($\mu$b/sr)\\
&&&\\
\hline
&&&\\
$-0.9$&$0.170\pm 0.029$&$-0.9$&$0.234\pm 0.027$\\
$-0.7$&$0.168\pm 0.033$&$-0.7$&$0.317\pm 0.040$\\
$-0.5$&$0.123\pm 0.029$&$-0.5$&$0.392\pm 0.049$\\
$-0.3$&$0.219\pm 0.039$&$-0.3$&$0.552\pm 0.061$\\
$-0.1$&$0.186\pm 0.036$&$-0.1$&$0.479\pm 0.057$\\
$\phantom{-}0.1$&$0.262\pm 0.044$&$\phantom{-}0.1$&$0.490\pm 0.057$\\
$\phantom{-}0.3$&$0.164\pm 0.032$&$\phantom{-}0.3$&$0.434\pm 0.060$\\
$\phantom{-}0.5$&$0.157\pm 0.031$&$\phantom{-}0.5$&$0.554\pm 0.069$\\
$\phantom{-}0.7$&$0.178\pm 0.034$&$\phantom{-}0.7$&$0.425\pm 0.061$\\
$\phantom{-}0.9$&$0.100\pm 0.024$&$\phantom{-}0.9$&$0.188\pm 0.038$\\
&&&\\
\hline
\end{tabular}
\end{table}

\begin{table}[hp]
\caption[table2]{Differential cross section with respect to the proton-proton
angle for the $pp\to pp\eta$ reaction at $Q=16$ and 37~MeV. In addition to the 
statistical error, there is an overall systematic uncertainty of about 20\%.}
\centering
\vspace{2ex}
\begin{tabular}{|c|c||c|c|}
\hline
\multicolumn{2}{|c||}{}&
\multicolumn{2}{c|}{}\\
\multicolumn{2}{|c||}{$Q =16$~MeV}&
\multicolumn{2}{c|}{$Q =37$~MeV}\\
\multicolumn{2}{|c||}{}&
\multicolumn{2}{c|}{}\\
\hline
&&&\\
$\cos\theta_{pp}$&$d\sigma/d\Omega_{pp}$ ($\mu$b/sr)&
$\cos\theta_{pp}$&$d\sigma/d\Omega_{pp}$ ($\mu$b/sr)\\
&&&\\
\hline
&&&\\
$0.05$&$0.196\pm 0.039$&$0.05$&$0.331\pm 0.043$\\
$0.15$&$0.150\pm 0.029$&$0.15$&$0.308\pm 0.041$\\
$0.25$&$0.166\pm 0.028$&$0.25$&$0.343\pm 0.043$\\
$0.35$&$0.182\pm 0.026$&$0.35$&$0.405\pm 0.047$\\
$0.45$&$0.169\pm 0.024$&$0.45$&$0.344\pm 0.042$\\
$0.55$&$0.190\pm 0.029$&$0.55$&$0.355\pm 0.042$\\
$0.65$&$0.125\pm 0.029$&$0.65$&$0.473\pm 0.050$\\
$0.75$&$0.163\pm 0.043$&$0.75$&$0.391\pm 0.050$\\
$0.85$&$0.169\pm 0.084$&$0.85$&$0.508\pm 0.070$\\
---&---&$0.95$&$0.88\pm 0.19$\\
&&&\\
\hline
\end{tabular}
\end{table}

\begin{table}[hp]
\caption[table3]{Differential cross section with respect to the $\eta$ c.m.\
kinetic energy for the $pp\to pp\eta$ reaction at $Q=16$ and 37~MeV. In 
addition to the statistical error, there is a typical systematic uncertainty 
of about 20\%, which increases at the upper end of the spectrum.}
\centering
\vspace{2ex}
\begin{tabular}{|c|c||c|c|}
\hline
\multicolumn{2}{|c||}{}&
\multicolumn{2}{c|}{}\\
\multicolumn{2}{|c||}{$Q =16$~MeV}&
\multicolumn{2}{c|}{$Q =37$~MeV}\\
\multicolumn{2}{|c||}{}&
\multicolumn{2}{c|}{}\\
\hline
&&&\\
$T_{\eta}$&$d\sigma/dT_{\eta}$ ($\mu$b/MeV)&
$T_{\eta}$&$d\sigma/dT_{\eta}$ ($\mu$b/MeV)\\
&&&\\
\hline
&&&\\
$ 0.5$&$0.046\pm 0.010$&$ 1.0$&$0.055\pm 0.010$\\
$ 1.5$&$0.074\pm 0.013$&$ 3.0$&$0.110\pm 0.014$\\
$ 2.5$&$0.089\pm 0.015$&$ 5.0$&$0.115\pm 0.015$\\
$ 3.5$&$0.114\pm 0.018$&$ 7.0$&$0.142\pm 0.017$\\
$ 4.5$&$0.114\pm 0.019$&$ 9.0$&$0.164\pm 0.020$\\
 $5.5$&$0.146\pm 0.025$&$11.0$&$0.175\pm 0.022$\\
 $6.5$&$0.173\pm 0.033$&$13.0$&$0.160\pm 0.021$\\
 $7.5$&$0.191\pm 0.045$&$15.0$&$0.244\pm 0.029$\\
 $8.5$&$0.059\pm 0.029$&$17.0$&$0.183\pm 0.028$\\
 $9.5$&$0.237\pm 0.090$&$19.0$&$0.155\pm 0.030$\\
$10.5$&$0.196\pm 0.098$&$21.0$&$0.216\pm 0.041$\\
 ---&--- &$23.0$&$0.214\pm 0.047$\\
 ---&--- &$25.0$&$0.212\pm 0.059$\\ 
 ---&--- &$27.0$&$0.160\pm 0.080$\\ 
&&&\\
\hline
\end{tabular}
\end{table}

\newpage
\noindent
{\bf\Large Figure Captions}\\[4ex]
Fig.~1. Differential cross section in the $\eta$-production angle for the
$pp\to pp\eta$ reaction at (a) $Q=37$~MeV ($T_p =1350$~MeV) and (b)
$Q=16$~MeV ($T_p=1295$~MeV). The 
dashed curves represent the Monte Carlo predictions of phase space modified by 
the proton-proton final state interaction, whereas the solid curve includes 
also the angular dependence of Eq.~(5) with $a=-15$ and $b=22$.\\[4ex]
Fig.~2. Differential cross section in the proton-proton production angle for the
$pp\to pp\eta$ reaction at (a) $Q=37$~MeV and (b) $16$~MeV with
curves as described in Fig.~1.\\[4ex]
Fig.~3. Distribution in $\eta$ kinetic energies from the $pp\to pp\eta$ reaction
at (a) $Q=37$~MeV and (b) $16$~MeV. The short-dashed curves represent the Monte
Carlo predictions of phase space and the long-dashed shows the influence of the
proton-proton final state interaction. The solid curve includes also the
modifications induced by the angular and momentum dependence of Eq.~(5) with 
$a=-15$ and $b=22$. Since the experimental acceptance changes somewhat 
according to the Monte Carlo generator used, the experimental points would be 
slightly lowered if we had extracted them using $a=b=0$.

\clearpage

\pagestyle{empty}
\begin{figure}
\includegraphics{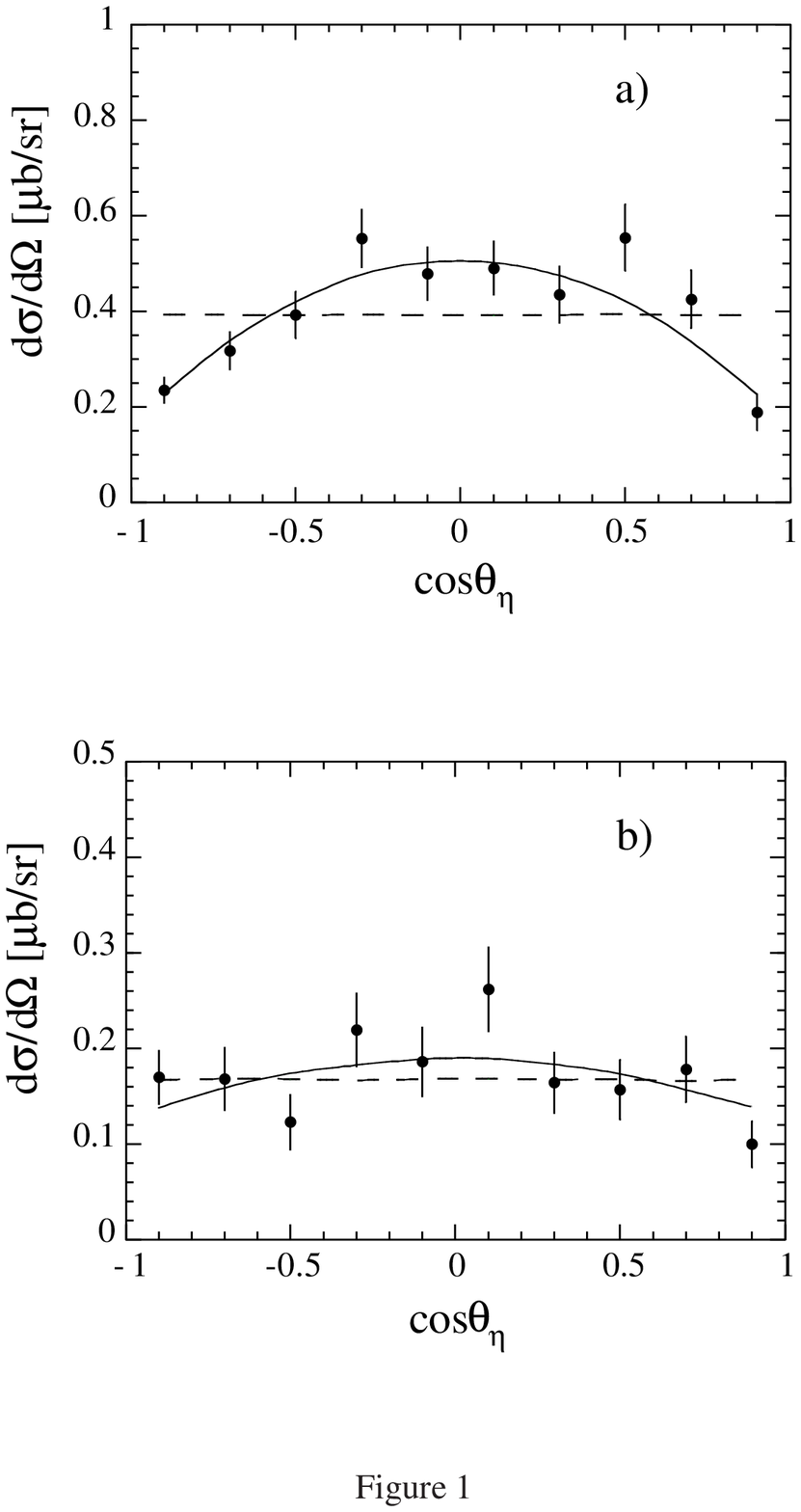}
\end{figure}

\setlength{\topmargin}{-20mm}
\begin{figure}
\includegraphics{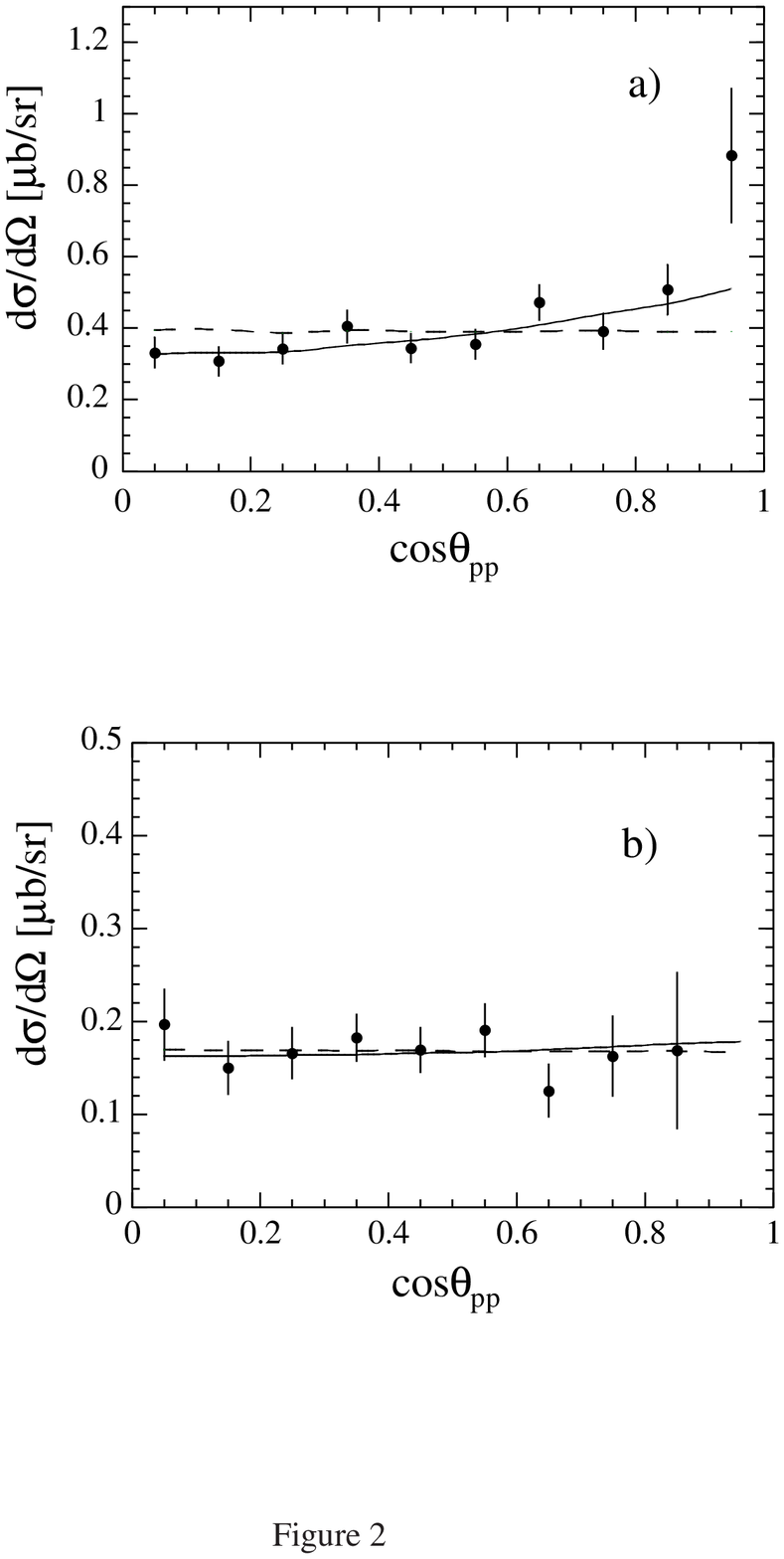}
\end{figure}

\setlength{\topmargin}{-20mm}
\begin{figure}
\includegraphics{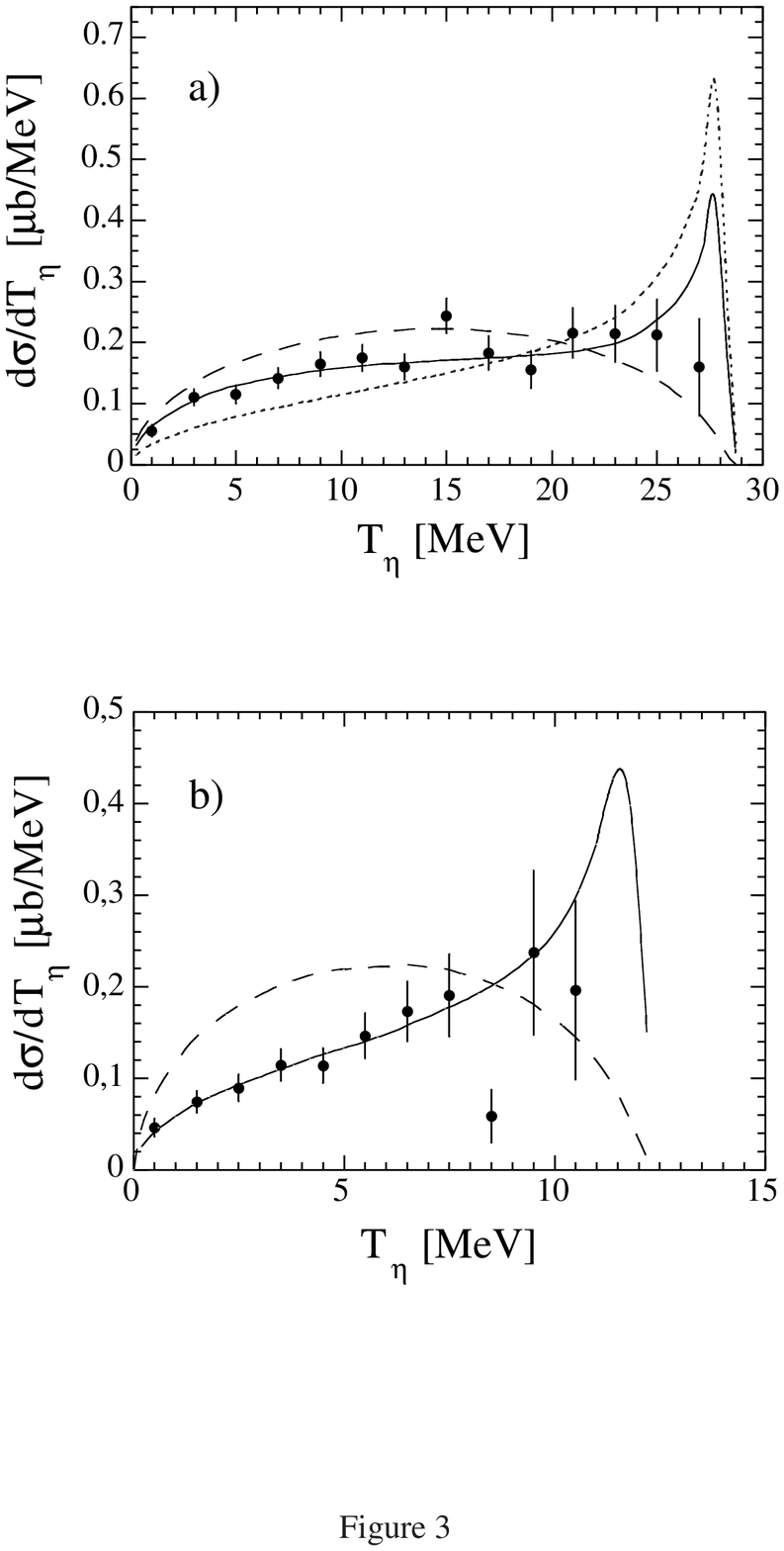}
\end{figure}

\end{document}